\documentclass[11pt,a4paper]{article}
\usepackage{jheppub}

\pdfoutput=1 

\usepackage[utf8]{inputenc}
\usepackage{amsmath}
\usepackage{xcolor}
\usepackage[english]{babel}
\usepackage{graphicx}

\newcommand{\ket}[1]{\left|#1\right\rangle}

\title{Black Hole Cannibalism}
\author[1]{Ning Bao}
\emailAdd{ningbao75@gmail.com}
\affiliation[1]{Computational Science Initiative, Brookhaven National
  Laboratory, Upton, New York, 11973.}
\affiliation[2]{Center for Theoretical Physics and Department of Physics,
     University of California, Berkeley, CA 94720.}
\author[2]{and Elizabeth Wildenhain}
\emailAdd{elizabeth\_wildenhain@berkeley.edu}
\date{August 2020}

\begin{document}

\abstract{
We consider a version of the Hayden-Preskill thought experiment in which the message thrown into the black hole is itself a smaller black hole. We then discuss the implications of the existence of a recovery channel for this black hole message at asymptotic infinity, resulting in a sharpening of the black hole information paradox for observers who never need to approach a horizon. We suggest decoherence mechanisms as a way of resolving this sharpened paradox.}

\maketitle    

\section{Introduction}

The firewall paradox \cite{Almheiri:2012rt, Marolf:2013dba} is a tension between four widely accepted postulates about black hole evaporation:
\begin{enumerate}
    \item \textit{Unitary}. There is a unitary governing the dynamics of black hole formation and evaporation.
    \item \textit{Semi-classical field theory}. Physics outside the stretched horizon of a large black hole is well approximated by semi-classical field equations of a low energy EFT with local Lorentz invariance.
    \item \textit{Black Hole Entropy/Discrete Energy Levels}. To distant observers, a black hole behaves like a quantum system with discrete energy levels, where the dimension of the subspace of states is the exponential of the Bekenstein-Hawking entropy.
    \item \textit{No Drama}. An observer who falls freely across the horizon experiences nothing unusual.
\end{enumerate}
Postulates (1) and (3) imply maximal entanglement between the late and early Hawking radiation, but postulates (2) and (4) imply that a portion of the late Hawking radiation is highly entangled with modes in the black hole interior. The full set of postulates, therefore, implies a violation of monogamy of entanglement for this portion of the late Hawking modes \cite{Almheiri:2012rt}.

Proposed resolutions include (i) modifying quantum mechanics \cite{Papadodimas:2012aq, Lloyd:2013bza}, (ii) allowing a breakdown of no drama \cite{Almheiri:2012rt}, (iii) violating unitarity \cite{Harlow:2014yka}, (iv) identifying earlier Hawking radiation with the black hole interior \cite{Maldacena:2013xja}, (v) modifying the interior geometry \cite{Mathur:2005zp,Nomura:2014woa, Hertog:2017vod}, (vi) invoking quantum complexity theory \cite{Harlow:2013tf,Bao:2016uan,Bao:2017who}, (vii) allowing for remnants \cite{Chen:2014jwq}, (viii) violating locality \cite{Giddings:2012gc,Osuga:2016htn}, and (ix) implementing an auxiliary system to collect the radiation \cite{Penington:2019npb, Almheiri:2019psf}. Many of these suggestions---in particular, options (ii), (iv), (v), (vi), and (vii)---appeal to the role of the horizon in effectively hiding a portion of the spacetime. In this essay we present a thought experiment that results in a manifest violation of monogamy of entanglement amongst modes at asymptotic infinity, in which none of the modes are hidden behind a horizon. This implies that hiding modes behind a horizon or appealing to strong gravitational effects are not sufficient resolutions. We argue that accounting for decoherence alleviates the tension.

\section{The Thought Experiment}

In broad strokes, our thought experiment consists of throwing a small black hole into a larger one and using the Hayden-Preskill protocol to recover the state of the smaller black hole. We argue that, given the above postulates and a qubit model of black holes, this yields a violation of monogamy of entanglement in states held only by the asymptotic observer.

Consider a black hole of mass $M_1$ formed from the collapse of matter in a pure state. As per Hawking's calculation \cite{Hawking:1974rv,Hawking:1976de}, the black hole will radiate maximally entangled pairs, one of which escapes the horizon and the other of which falls into the black hole interior. We model this black hole as a collection of $N_1$ qubits at its formation and its evaporation as the release of individual qubits into the environment \cite{Hayden:2007cs}. The interior qubits undergo scrambling dynamics. We can factorize the total Hilbert space as: 
\begin{equation}
    \mathcal{H}_1=\mathcal{H}_{\mathrm{BH}_1}(t)\otimes\mathcal{H}_{\mathrm{rad}_1}(t),
\end{equation}
where $\mathcal{H}_{\mathrm{BH}_1}(t)$ corresponds to the Hilbert space of the remaining black hole and $\mathcal{H}_{\mathrm{rad}_1}(t)$ denotes that of the Hawking radiation after evolution for some time $t$.\footnote{Although this evolution is truly time evolution, the choice of time coordinate is arbitrary \cite{Bao:2017who}. In addition, one can use finite dimensional quantum information theory to describe black hole dynamics \cite{Bao_2017}.} The Hilbert space factors are time-dependent because qubits are transferred from the black hole to the radiation as the black hole evaporates. According to unitarity, the total state $\ket{\psi_1(t)}$ is pure at all times. Let us denote the state of the black hole and its Hawking radiation emitted up to time $t$ as $\rho_{\mathrm{BH}_1}(t)$ and $\rho_{\mathrm{rad}_1}(t)$ respectively, each defined via a partial trace over the neglected system. Assume that the asymptotic observer has been collecting the radiation from this black hole since its formation.

\begin{figure*}
    \centering
    \includegraphics[width=\textwidth]{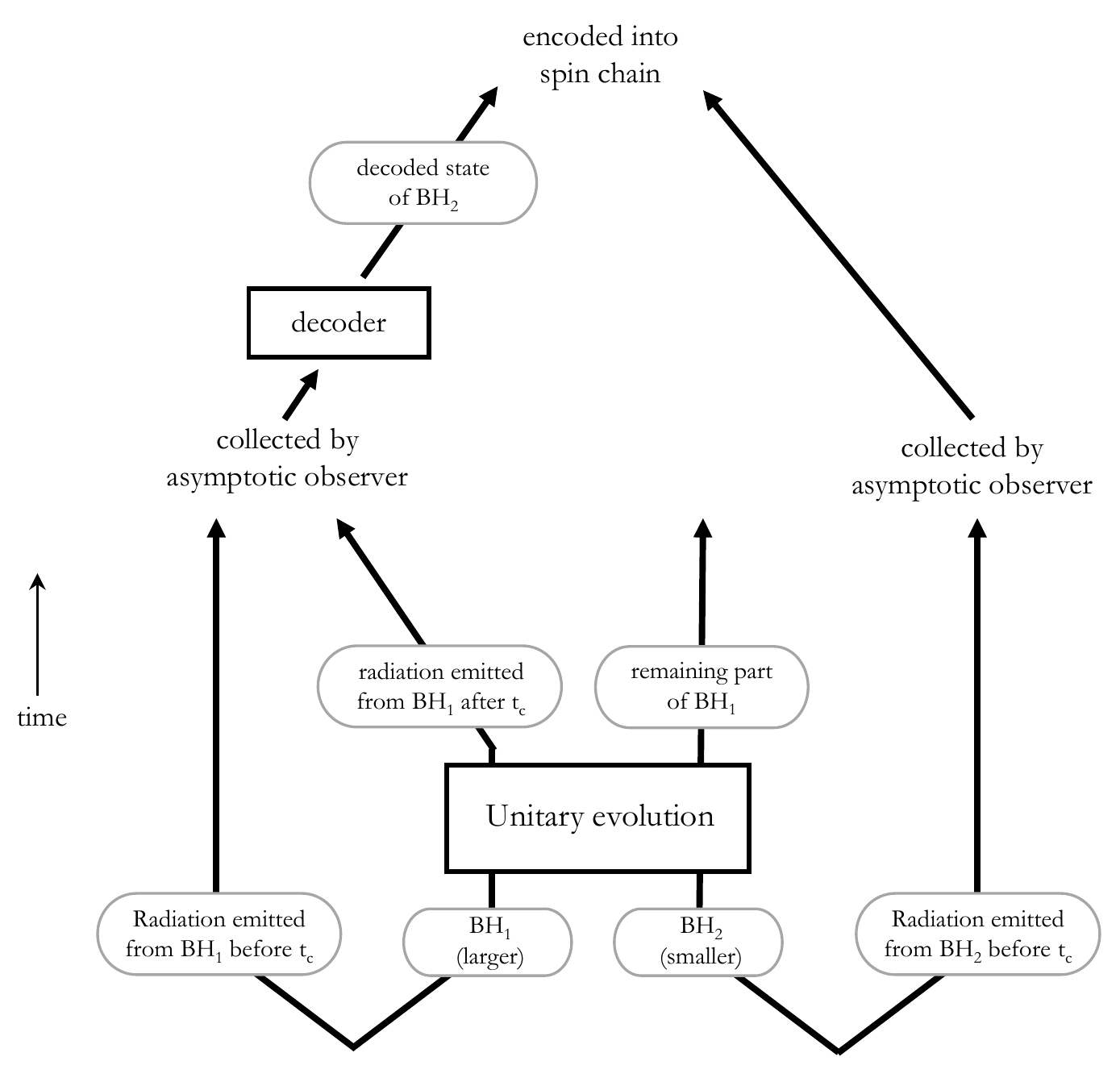}
    \caption{A diagram of the Hayden-Preskill protocol in which the ``message" is another black hole. Black hole 1 (BH$_1$), the larger black hole, has been evaporating for a long time and has become maximally entangled with its previously emitted radiation. The ``message" is the state of black hole 2 (BH$_2$), the smaller black hole, which is also maximally entangled with its previously emitted radiation. The message (BH$_2$) falls into BH$_1$ and the joint system evolves unitarily. More radiation is emitted from BH$_1$ and collected by the asymptotic observer. The asymptotic observer runs this radiation and the radiation previously emitted from BH$_1$ through a decoder to recover the state of BH$_2$. The asymptotic observer also collects the radiation emitted from BH$_2$ before it fell into BH$_1$.}
    \label{fig:2}
\end{figure*}

Now suppose there is another, smaller black hole in this spacetime with mass $M_2<<M_1$, again modeled as a collection of $N_2$ qubits at its formation, evaporating via release of qubits. Assume that the asymptotic observer collects the radiation from this black hole until, at some time $t_c$ after its Page time, it falls into BH$_1$.\footnote{One may be concerned that some radiation from BH$_1$ may fall into BH$_2$, preventing it from being collected. We can avoid this by placing the black holes in separate regions connected by an (initially non)traversable wormhole. We can choose the momenta of the black holes and when the wormhole is made traversable such that the black holes collide at the appropriate time. To avoid complications with emission of radiation during transit through the wormhole, we can tune the wormhole to be very short, so that the transit time is much smaller than the expected time elapsed between emission of Hawking quanta. Alternatively, we could construct large Dyson spheres surrounding the black holes, which collect the HR from each of them until their point of collision.} The observer can identify which modes correspond to which black hole by measuring the temperature.\footnote{Or, the modes can be easily distinguished because the BH's have not yet been sent through the traversable wormholes and are therefore separate. Or if using Dyson spheres, the observer can see which Dyson sphere collected which mode.} After BH$_2$ has fallen into BH$_1$, the asymptotic observer continues collecting radiation emitted by BH$_1$. See figure \ref{fig:2} for a summary of this sequence of events.

Hayden and Preskill demonstrated that it is information-theoretically possible for an observer outside a black hole to reconstruct the state of a quantum system (the``message") thrown into the black hole \cite{Hayden:2007cs}. The observer first collects the Hawking radiation until they possess a system maximally entangled with the black hole, which occurs at the Page time. The message is then thrown into the black hole, which evolves via a deterministic unitary transformation that thoroughly and quickly mixes the message into the black hole's state. As the black hole continues to evaporate, the observer collects the radiation. The outside observer needs to collect only a few more qubits than the length of the original message to reconstruct it.

We choose that the state of the radiation released by BH$_1$ is maximally entangled with BH$_1$ at $t_c$. Our asymptotic observer therefore has the ingredients to perform the Hayden-Preskill protocol to recover the state of BH$_2$ when it fell into BH$_1$, $\rho_{\mathrm{BH}_2}(t_c)$. Recall that the asymptotic observer also collected the radiation from BH$_2$, which we can partition into ``early" and ``late" radiation corresponding to emission before and after the Page time of BH$_2$ respectively. Therefore, the asymptotic observer possesses the following states:
\begin{enumerate}
    \item $\rho_{\mathrm{earlyRad}_2}(t_c)$, containing $N_2/2 + \delta$ qubits and defined from $\rho_{\mathrm{rad}_2}(t_c)$ by tracing out the qubits associated with the late radiation.
    \item $\rho_{\mathrm{lateRad}_2}(t_c)$, defined from $\rho_{\mathrm{rad}_2}(t_c)$ by tracing out the qubits associated with the early radiation.
    \item $\rho_{\mathrm{BH}_2}(t_c)$, reconstructed via the Hayden-Preskill protocol.
\end{enumerate}
See figure \ref{fig:2} for a circuit-style diagram of this scenario. 

\begin{figure*}
    \centering
    \includegraphics[width=.9\textwidth]{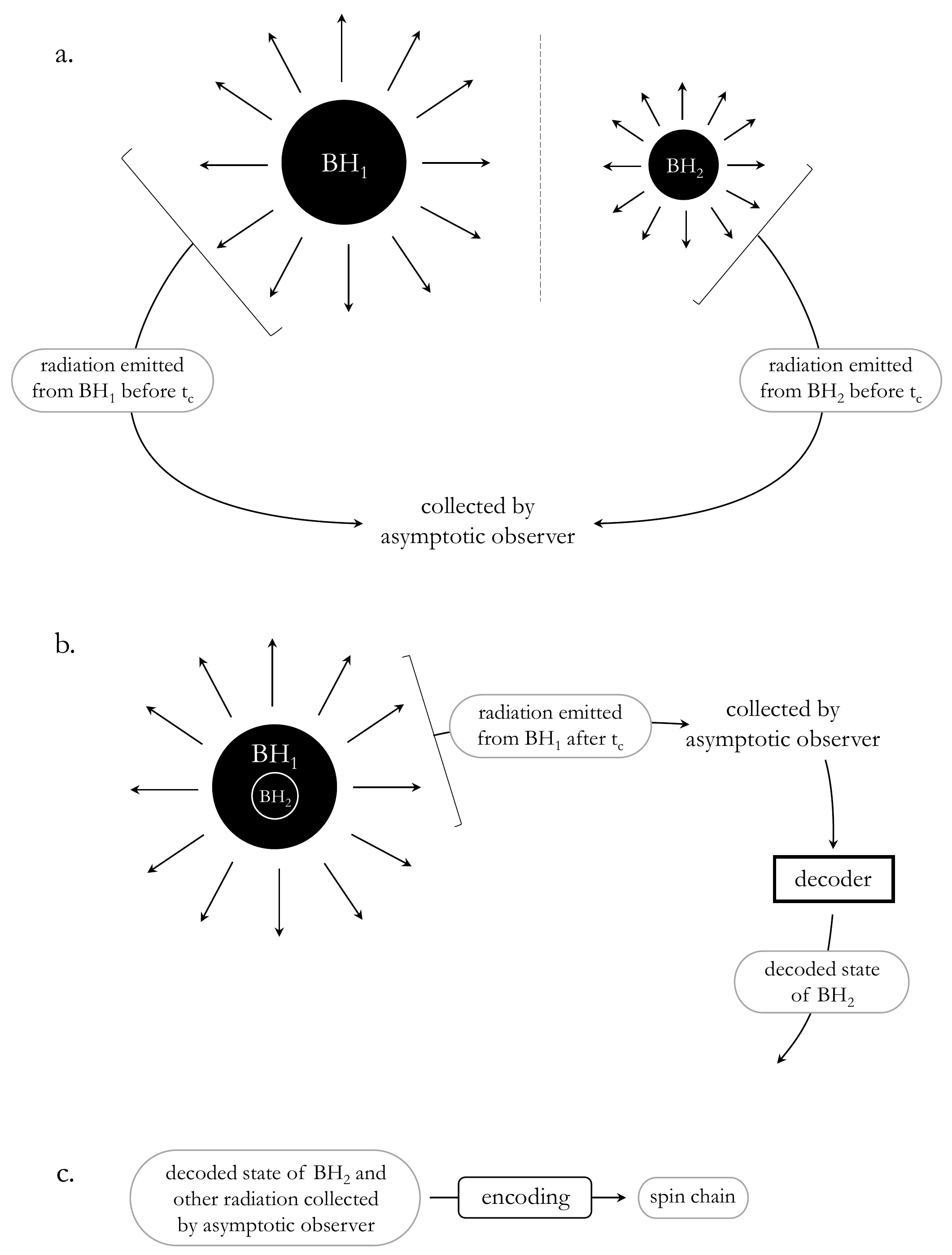}
    \caption{The three main steps of the thought experiment. \textit{a}. The two black holes are initially in separate regions (see footnote 2), emitting Hawking radiation. The asymptotic observer collects the radiation emitted from both black holes. \textit{b}. The BH$_2$ has fallen into BH$_1$. BH$_1$ continues to emit radiation, which is collected by the asymptotic observer. The observer runs this radiation (and the radiation previously collected from BH$_1$) through a decoder. \textit{c.} Finally, all the radiation in the possession of the asymptotic observer (including the reconstructed state of BH$_2$) is encoded into a spin chain.}
    \label{fig:1}
\end{figure*}

As we shall see, these systems exhibit a violation of monogamy of entanglement. Page's theorem \cite{Page:1993df} states that if we bipartition a statistically typical\footnote{With respect to the Haar measure. For a more complete discussion, see \cite{Hayden:2007cs,Bao:2017gza,Sekino:2008he}.} quantum system into $A$ and $B$ with Hilbert space dimensions $m$ and $n$ respectively, and if $1\ll m \leq n$, then 
\begin{equation}
    S_{m,n} \approx \ln m - \frac{m}{2n}.
\end{equation}
This implies that if the joint system $AB$ is pure and if $m$ and $n$ are very large, then the smaller subsystem is close to maximally mixed. In our setup, the qubits of (2) and (3) are $A$ and the qubits of (1) are $B$ of a pure joint system $AB$. We can apply Page's theorem by choosing appropriate system sizes. Therefore, the qubits making up subsystems (2) and (3) are maximally entangled with (1). A maximally mixed state has a diagonal density matrix, so $\rho_{2,3}$ is roughly diagonal. Tracing out the qubits of (3) thus yields a smaller-dimensional nearly diagonal density matrix for the qubits of (2). Therefore, (2) is nearly maximally entangled with (1). 

Just before BH$_1$ swallows BH$_2$, Hawking's calculation implies a high degree of entanglement across the horizon of BH$_2$. Therefore, the qubits identified with the BH$_2$'s interior and those identified with its radiation are highly entangled. Crossing the event horizon of BH$_1$ should not change the entanglement structure of BH$_2$'s interior. Because the recovery procedure involves only LOCC operations, the entanglement structure of (3) is the same as that of the original state. Therefore, subsystem (3) is highly entangled with (2). Because (2) is maximally entangled with (1), this is a violation of monogamy of entanglement. Therefore, the 3 subsystems held by the asymptotic observer exhibit a violation of monogamy of entanglement.

This thought experiment thus presents a sharp version of the firewall paradox for the \textit{asymptotic observer}, who is free to manipulate these states, e.g. by encoding them in a spin chain, and has no time constraint on doing so. Therefore, no resolution to this paradox can rely on strong gravitational effects or on hiding something behind the black hole horizon.

\section{A Proposed Resolution via Decoherence}

In the spirit of \cite{Bao:2017who, Bao:2020zdo}, we argue that accounting for decoherence resolves this paradox. Decoherence is the observation that quantum systems interacting with an environment become entangled with or, equivalently, leak information into said environment \cite{Zeh:1970zz,Zurek:1981xq,Griffiths:1984rx,Joos:1984uk}. Decoherence causes the system's the density matrix to become nearly diagonal in a special basis, which is determined by details of the interaction. Frequently paired with decoherence is the Everettian view of quantum mechanics, which holds that measurement restricts the observer to a particular ``branch" of the wavefunction corresponding to a measurement outcome.\footnote{The distinction between this view and wavefunction collapse is philosophical---the Everettian view asserts that the global wavefunction continues to exist post-measurement, while collapse-based interpretations deny this. Thus in the Everettian view, all measurement possibilities are realized within the global wavefunction as separate branches, hence the alternative name ``Many Worlds interpretation." There is actually a family of related views, many of which are called the Everettian interpretation or the Many Worlds interpretation. Sometimes these terms refer to distinct views in this family, and other times they are conflated. The subtleties involved are not critical for our discussion.} The Everettian picture may seem to privilege an arbitrary basis for the decomposition of the global wavefunction into branches, but decoherence provides a particular basis in which the density matrix becomes essentially diagonal, corresponding to a superposition of classical macroscopic states.

For evaporating black holes, wavefunction branching occurs when something interacts with its released Hawking quanta, which have, for example, a range of possible outgoing momenta. A ``definite geometry" requires a choice of momentum for each emitted Hawking mode and, by extension, a definite position and momentum for the black hole. The total state at non-asymptotic values of $t$ describes an ensemble of these possible geometries \cite{Bao:2017who}. A classical (definite) geometry exists only on a branch of the wavefunction. As the Hawking modes experience interactions, the state of the black hole decoheres into some preferred basis. 

Fully defining the process by which BH$_2$ falls into BH$_1$ requires a definite point and time at which BH$_2$ crosses the event horizon of BH$_1$, thus requiring a definite geometry, or perhaps a projection to a subset of geometries. Therefore, an instance of this process must occur on a branch (or subset of branches) rather than on the global wavefunction. The asymptotic observer in our setup, however, sees the global wavefunction, which is a superposition of geometries. Therefore, the global observer cannot possess a definite reconstructed state $\rho_{\mathrm{BH}_2}(t_c)$, meaning the global observer does not possess definite states in violation of monogamy of entanglement. Thus, accounting for decoherence resolves the paradox we have outlined in this work.\footnote{Our proposed resolution is related to recent works applying the RT formula to a black hole coupled to an auxiliary system \cite{Penington:2019npb, Almheiri:2019psf}. The lesson from decoherence for the firewall paradox is that if the bulk observer has access only to the system, their experiencing information loss does not conflict with unitary evolution of the system + environment. The recent works applying the RT formula to a black hole + auxiliary system have the same form: the ``environment" is the auxiliary system and the black hole is the ``system." Escape of Hawking radiation in \cite{Penington:2019npb} with absorbing boundary conditions corresponds to leakage of information into the environment, and the bulk observer has access only to the system. Again, the lesson is that confusing a part (the BH) with the whole (BH + auxiliary system) leads to a contradiction.}

\begin{acknowledgments}
We would like to thank Aidan Chatwin-Davies for useful comments on the draft. We also thank the attendees of the GeoFlow January 2021 Collaboration Meeting for helpful feedback. N.B. is supported by the National Science Foundation under grant number 82248-13067-44-PHPXH, by the Department of Energy under grant number DE-SC0019380, and by the Computational Science Initiative at Brookhaven National Laboratory. E.W. is supported by the Berkeley Connect Fellowship.

\end{acknowledgments}

\bibliographystyle{unsrt}
\bibliography{BHHP}

\end{document}